\documentclass{aastex}
\usepackage{spr-astr-addons}
\usepackage{url}
\usepackage{graphicx}
\usepackage{array}
\usepackage{makecell}

\RequirePackage{color}

\begin{document}

\title{Characteristics of events with metric-to-decahectometric type II radio bursts associated with CMEs and flares in relation to SEP events}
\slugcomment{Not to appear in Nonlearned J., 45.}
\shorttitle{Characteristics of m-to-DH type II radio bursts}
\shortauthors{Prakash et al.2017}

\author{O.Prakash \altaffilmark{1}} \and \author{Li Feng\altaffilmark{1}} \and \author{G. Michalek\altaffilmark{2}} \and \author{Weiqun Gan\altaffilmark{1}} \author{Lei Lu \altaffilmark{1,3}} \and \author{A. Shanmugaraju \altaffilmark{4}} \and \author{ S. Umapathy \altaffilmark{5}}


\altaffiltext{1}{Key Laboratory of Dark Matter and Space Astronomy, Purple Mountain Observatory, Chinese Academy of Sciences, Nanjing -210008, Jiangsu, China.}
\altaffiltext{2}{Astronomical Observatory of Jagiellonian University, Krakow, Poland.}
\altaffiltext{3}{University of Chinese Academy of Sciences, Beijing – 100049, China.}
\altaffiltext{4}{Department of Physics, Arul Anandar College, Karumathur, Madurai- 625514, Tamil Nadu, India.}
\altaffiltext{5}{School of Physics, Madurai Kamaraj University, Madurai - 625021, Tamil Nadu, India.}

\begin{abstract}
A gradual solar energetic particle (SEP) event is thought to happen when particles are accelerated at a shock due to a fast coronal mass ejection (CME). To quantify what kind of solar eruptions can result in such SEP events, we have conducted detailed investigations on the characteristics of CMEs, solar flares and metric-to-decahectometric wavelength type II radio bursts (herein after m-to-DH type II bursts) for SEP-associated and non-SEP-associated events, observed during the period of 1997--2012. Interestingly, 65\% of m-to-DH type II bursts associated with CMEs and flares produced SEP events. The SEP-associated CMEs have higher sky-plane mean speed, projection corrected speed, and sky-plane peak speed than those of non-SEP-associated CMEs respectively by 30\%, 39\%, and 25\%, even though the two sets of CMEs achieved their sky-plane peak speeds at nearly similar heights within LASCO field of view. We found Pearson's correlation coefficients between the speeds of CMEs (sky-plane speed and corrected speed) and logarithmic peak intensity of SEP events are \emph{cc} = 0.62 and \emph{cc} = 0.58, respectively. We also found that the SEP-associated CMEs are on average of three times more decelerated (--21.52 m s$^{-2}$) than the non-SEP-associated CMEs (-- 5.63 m s$^{-2}$ ). The SEP-associated flares have a mean peak flux (1.85 $\times$ 10$^{-4}$ W m$^{-2}$) three times larger than that of non-SEP-associated flares, even though the flare duration (rise time) of both sets of events is similar. The SEP-associated m type II bursts have higher frequency drift rate and associated shock speed than those of the non-SEP-associated events by 70\% and 25\% respectively. The average formation heights of m and DH type II radio bursts for SEP-associated events (1.31 \emph{R}$_\circ$ and 3.54 \emph{R}$_\circ$, respectively) are lower than for non-SEP-associated events (1.61 \emph{R}$_\circ$ and 3.91 \emph{R}$_\circ$, respectively). 93\% of SEP-associated events originate from the western hemisphere and 65\% of SEP-associated events are associated with interacting CMEs. The obtained results indicate that, at least for the set of CMEs associated with m-to-DH type II bursts, SEP-associated CMEs are more energetic than those not associated with SEPs, thus suggesting that they are effective particle accelerators.
\end{abstract}


\section{Introduction}
 ~~ Solar energetic particle (SEP) events are thought to be originated from solar flares and/or coronal mass ejections (CMEs) \citep{Reames1999,Reames2013}. In general, these particle events are classified into two categories: impulsive SEP events typically last a few hours, and are closely associated with solar flares\citep{Anastas2002,Klein2005}; while large gradual events typically last a few days, and are associated with CME-driven shocks \citep{Reames1999,Kahler2001,Cane2006}. Furthermore, large fraction of SEP-associated CMEs is closely related with solar flares. Hence, some recent studies were demonstrated to add one more category, that is hybrid or mixed events in which, both solar flares and CME-driven shocks accelerate the particles that contribute to large SEP events \citep{Kocharov2002,Kallenrode2003,Gopal2001b}. These hybrid SEP events seem to be gradual SEP events but have impulsive-like properties(Vainio \emph{et al.,} 2007).\\
\indent Several authors have pointed out that the importance to study the SEP events associated with solar flares and CMEs (hybrid/mixed events). For example, \cite* {Gopal2001a} demonstrated that interaction with pre-CMEs was a powerful discriminator for predicting SEP events, in addition to other properties of primary CMEs. They also suggested that type II associated CMEs are good electron and proton accelerators. But, \cite*{Kahler2014} concluded that the higher SEP peak intensity values with pre-CMEs may not be primarily due to CME interactions, such as the "twin-CME" scenario, but they were explained by a general increase of both background seed particles and more frequent CMEs during solar maximum. Therefore, the relevance of CME interactions for larger SEP event intensities remains unclear \citep{Gopal2003,LugaFar2014,Kahler2014}. \cite*{Gopal2014} reported three key factors affecting the occurrence of large SEPs and ground level enhancement (GLE) events: CME speed, magnetic connection, and ambient solar wind properties. On the other hand, \cite*{Shan2014} investigated 25 interacting CMEs and their associated flare and SEP activities. They found that the primary CMEs were more energetic than the pre-CMEs, and their associated X-ray flares were also stronger. The SEP intensity was positively related to the integrated flux of X-ray flares associated with the primary CMEs. In another study, \cite*{Trottet2015} carried out a statistical analysis on the contributions of flares and CMEs to major SEP events. They showed that the only parameters that significantly affect the SEP intensity are the CME speed, and SXR fluence. Thus solar flare associated fast and wide CME-driven shocks (hybrid events) can cause large SEP events. These hybrid events may play an important role in determining the space-weather conditions (Bothmer and Daglis, 2007).\\
\indent The type II radio bursts observed in coronal (in metric wavelength) and interplanetary (IP) medium (in Deca-Hectometric wavelength) are considered to be produced by the acceleration of non-thermal electrons. They occur at the local plasma frequency and/or its harmonic. By assuming a coronal/IP density model, the frequency of emission can be converted into the heliocentric distance at which the radio emission originates \citep{PayScot1947,WildMc1950}. Radio emission can be observed in the spectral domain of metric (m) or Deca-Hectometric (DH) or kilometric (km) or in any of the two domains or in the entire radio spectral domain (herein after we refer to them as m-to-km type II bursts). Shocks are formed when CME speed exceeds Alfven speed. A close relationship between kinetic energy of CMEs and the wavelength range of associated type II radio bursts has been reported by a few authors \citep{Gopal2001a,Gopal2005,Lara2003,Prakash2014}. These CMEs have been found to be proton and electron accelerators, and are capable of producing various space weather conditions (Bothmer and Daglis, 2007). They can also cause geomagnetic storms when they are Earth-directed.\\
\indent The m type II radio bursts occur in the frequency range of 150 – 15 MHz (Nelson and Melrose, 1985) and the shocks excited in the distance range 1-3 \emph{R}$_\circ$ (1\emph{R}$_\circ$ = 6.96 $\times$ 10$^{5}$ km). The lower limit 15 MHz is the ionospheric cutoff frequency. On the other hand,  DH type II radio bursts originate roughly in the outer corona within the heliocentric distance of about 2–10 Ro. DH type II radio bursts are also observed as a continuation of m type II radio bursts (herein after m-to-DH type II radio bursts). DH and km type II radio bursts are recorded in the frequency ranges 14–1 MHz and 1 MHz – 20 kHz by RAD2 and RAD1 instruments respectively. These instruments are part of the Radio and Plasma Wave (\emph{WAVES}) experiment \citep{Bougeret1995} on board the Wind spacecraft \citep{Acu1995}.\\
\indent The characteristics of faster and wider CMEs associated with type II bursts make them different from that of average CMEs \citep{Gopal2005,Gopal2001a}. The kinetic energy of the CME determines the occurrence of the type II burst in different spectral domain \citep{Gopal2005}. \cite*{Gopal2005}studied the hierarchical relationship among type II bursts in different wavelength domains during the period 1997 - 2004. Where, they also found that the majority (78\%) of the m-to-km type II bursts were associated with SEP events. The solar sources of the small fraction of m-to-km type II bursts without SEP association were found to be due to poor connection to the near-Earth observer. Finally, they showed that the m-to-km type II bursts were associated with bigger flares compared to the purely m type II bursts.  The m-to-km type II radio bursts are associated with more energetic CMEs than that of type II radio burst observed in any single spectral domain. In addition, recent studies have pointed out that CMEs associated with longer and stronger solar flares can generate type II radio bursts in different radio wavelength domains. It could be an early indicator to predict the space weather conditions \citep{Michalek2003,Yashiro2004,Gopal2005,Prakash2009,Prakash2010,Prakash2014,Suresh2015} .\\
\indent Recently, \cite*{Cane2010} studied a set of 280 solar proton events that extended above 25 MeV and occurred in the years 1997-2006. They divided the events into five groups based on the relative abundances and particle profiles. They found that the least intense and relatively short-lived proton events that are electron-rich have association with type III radio bursts that occur at the start of the flare.  For an another case,\cite*{Papaioannou2016}studied the characteristics of a large data set of 314 hybrid events during the period 1984 – 2013. They found that most of the SEP events in their catalogue do not conform to a simple two categories and also they demonstrated that the velocity of the CME and flare peak flux are significantly correlated with proton peak flux and fluence in $>$10 MeV energy channel. Furthermore, there have been a wealth of statistical studies of SEP events \citep{Belov2005,Cane2010,Dierc2015,Papaioannou2016}, several of which have divided solar eruptions into SEP-associated and non-SEP associated events. However in this present paper, we have particularly made the list of events in which DH type II radio bursts is continuation of m type II bursts (m-to-DH type II bursts) associated with hybrid SEP events and divided these solar eruptions into SEP-associated and non-SEP-associated events. Further, we tried to quantify the differences between both sets of events during the Solar Cycle 23 and 24 (SC 23 and SC 24). The various properties of CMEs, solar flares and m and DH type II radio bursts associated with both sets of events are extensively compared in the present study. We also examined the relative timing of m and DH type II bursts, flares and CMEs for both sets of events. This kind of study is important to predict the relationship among CMEs, solar flares and type II radio bursts associated with SEP events. In Section 2, we describe the method of event selection and data analysis and in Section 3, results and discussions are given. A brief summary and conclusions are presented in Section 4.\\
\section{Event Selection and Analysis}
\indent In this paper, we define a SEP event according to its peak proton flux intensity in $>$10 MeV channel which should be larger than 1 \emph{pfu} (1 \emph{pfu} = 1 particle cm$^{-2}$ s$^{-1}$ sr$^{-1}$).  The method of searching m-to-DH type II radio bursts, the key parameters of the type II bursts and their associated flares and CMEs will be presented in the following paragraphs.\\
\indent We have considered a large sample of 400 DH type II radio bursts recorded during the period April 1997 – October 2012, which are listed in the online '\emph{CDAW Wind/WAVES}' type II catalog\footnote{$http://cdaw.gsfc.nasa.gov/CME_list/radio/waves_type2.html.$}. To identify the sample of m-to-DH type II bursts events in which DH type II bursts are a continuation of m type II bursts, we have compiled the corresponding m type II data observed by the \emph{Culgoora radio spectrograph}\footnote{$http://www.sws.bom.gov.au/World_Data_Centre/1/9.$} and further we followed the data selection criteria as presented in \cite*{Prakash2009}: \emph{i)} time delay between the DH type II radio bursts and m type II radio bursts should be less than an hour; \emph{ii)} the starting/ending frequencies of m type II radio bursts should be certain. Frequency-time drift graph, using the starting and ending frequencies and times of both m and DH type II bursts, is drawn for each event. Using these drift graphs, we have established a set of 48 m-to-DH type II bursts along with their associated solar flares and/or CMEs for our further analysis. Please note that the associations with flares and CMEs considered were adopted from the CDAW Wind/WAVES type II catalog. The starting and ending frequencies and times of m type II bursts were adopted from Culgoora event summary$^{2}$. These values of DH type II bursts are obtained from the \emph{CDAW Wind/WAVES} catalog$^{1}$
\begin{figure}[h]
\begin{center}
  \includegraphics[width=2.8in]{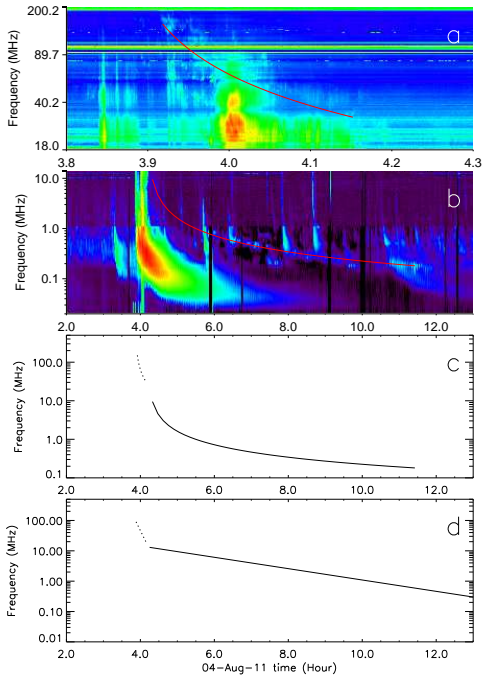}
 \caption{An example of the frequency-time drift graphs of the type II burst on 2011-August-04 in the m-to-DH type II domain: (a) the dynamic spectrum observed by Culgoora radio spectrograph; (b) the dynamic spectrum observed by Wind/WAVES. The red curves are the traced harmonic component of the type II bursts observed by Culgoora and Wind/WAVES, respectively. For the details of the tracing process, please check the corresponding text; (c) the DH type II extension of the m type II burst observed in m and DH-to-km domains, the traced harmonic emission in both domains are indicated by the dashed line and the solid line, respectively; (d) frequency-time drift graph based on the Prakash et al. (2009). The starting and ending frequencies and times in m and DH emission ranges were adopted from Culgoora event summary2 and CDAW Wind/WAVES catalog1 respectively.}
\label{Fig1}
\end{center}
\end{figure}

\indent In order to illustrate an example of a m-to-DH type II radio burst event, we include (Figure 1a-d) the frequency-time drift graphs on 2011-August-04. Figure1a shows the trace of second harmonic component of the type II emission over-plotted on the dynamic spectrum observed by the Culgoora radio spectrograph in the m domain. To make the tracing of the type II emission in the frequency-time $(f-t)$ frame, we first pinpointed the second harmonic emission along a straight line in the $1/f-t$ frame. Then the traced points were linearly fitted to produce the smooth curve in the $(f-t)$ frame. Similarly, the smooth curve in Figure 1b is the traced second harmonic component over-plotted in the dynamic spectrum observed by \emph{Wind/WAVES} in the DH domain. To prove that the traced type II emissions in Figure 1a and 1b are from the same event, we plot them in the same figure as shown in Figure 1c. The type II emissions in m and DH domains are indicated by dashed and solid lines, respectively. We can see in this figure that besides the time continuation, the slopes $(df/dt)$ of the two traced branches have a smooth transition from m to DH domains around 04:12 UT. Furthermore, the drift graph for m-to-DH type II burst based on \cite*{Prakash2009} is shown in Figure 1d.\\
\indent The start, peak, and end times (not listed in \emph{Wind/WAVES} catalog) of the flares are taken from the \emph{GOES} (Geosynchronous Operating Environmental Satellite) soft X-ray data\footnote{$ftp://ftp.ngdc.noaa.gov/STP/space-weather/solar-data/solar-features/solar-flares/x-rays/goes/.$}  by checking the source location of the CMEs and the corresponding flare class. The CME onset time ( onset time reported in the catalog is estimated by back-extrapolating the CME height–time data to 1 \emph{R}$_\circ$ from the constant-speed approximation method), first appearance time (it represents the time of CME at its first appearance in the \emph{LASCO} C2/C3 field of view), residual acceleration (obtained from a quadratic fit to the height–time measurements within the \emph{LASCO} FOV) of CMEs\footnote{$http://cdaw.gsfc.nasa.gov/CME_list/index.html.$}  are also obtained from the \emph{SOHO/LASCO} (Solar and Heliospheric Observatory/Large Angle and Spectrometric Coronagraph) catalog, maintained by the \emph{NASA's CDAW} data Center \citep{Yashiro2004,Gopal2009a}.
\begin{figure}[h]
\begin{center}
  \includegraphics[width=2.8in]{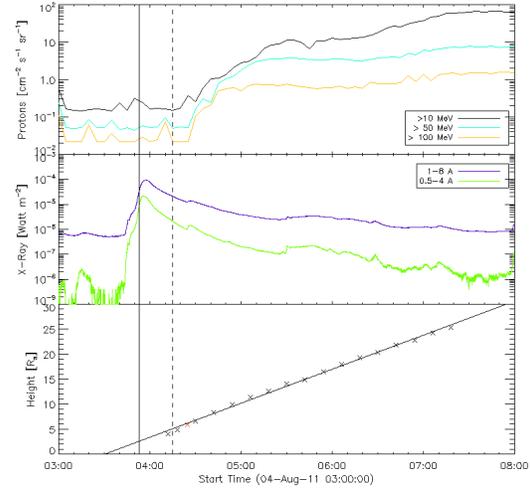}
  \caption{Top and middle panels give the SEP proton flux observed by GOES in three different energy channels (10, 50 and 100 MeV) and associated soft X-ray GOES light curves, respectively. Height-time measurement of SEP-associated CMEs on 2011-August-04 is shown in bottom panel. The start time of m and DH type II burst is marked as solid and dotted vertical lines, respectively. The red color height-time data point (please note these data points were obtained from the CDAW SOHO/LASCO CME catalog) in the bottom panel represents the sky-plane peak speed achieved by the SEP-associated CME within the LASCO FOV. }
\label{Fig2}
\end{center}
\end{figure}

\indent For each event, we plotted and visually examined the proton flux ($>$10 MeV), and also conformed their spatial and temporal association from CDAW major SEP events list\footnote{$https://cdaw.gsfc.nasa.gov/CME_list/sepe/.$}. Out of 48 events, 31 (24 major (\emph{Ip} $\geq$ 10) and 7 minor SEP events (1\emph{pfu} $\leq$ \emph{Ip} $\leq$ 10 \emph{pfu})) m-to-DH type II radio bursts were found to associate with SEP events and the remaining 17 of them were not associated with SEP events. As an example, SEP proton flux, \emph{GOES} X-ray light curves, and CME height-time diagram for one of the 31 selected events associated with SEPs, are shown in Figure 2.  This SEP-associated CME (with an average sky-plane speed of 1315 km s$^{-1}$) responsible for the m-to-DH type II burst was a full halo CME ejected from source location N19W36 close to the disk center. It was first observed by the C2 coronagraph in \emph{LASCO} at a height of 4.08 \emph{R}$_\circ$ at 04:12 UT. It was more decelerated due to the aerodynamic drag with a mean residual deceleration of --41.1 m s$^{-2}$ within the LASCO Field of view (FOV).The height where it achieved the sky-plane peak speed (2069 km s$^{-1}$) was nearly 5.96 \emph{R}$_\circ$.  It is shown in red colored data point in the height-time plot (bottom panel of Figure 2). This CME was also associated with a gradual M class solar flare (M9.3) with a duration of 23 min. After nearly 18 min of first appearance of the CME, the particle enhancement was observed by the GOES in the energy channel $>$ 10 MeV at 04:30 UT and it reached the peak intensity of 90 pfu at 21: 50 UT on 2011-August-05.The m and DH type II bursts formed at 03:53 UT and 04:15 UT, respectively during the rising and decaying phases of the solar flare. The particle acceleration happened only after the DH type II emission from the outer corona. For convenience of descriptions, hereinafter, a CME, flare, m and DH type II radio burst associated with/without a SEP event are called as SEP-associated/non-SEP-associated events.\\
\indent In Table 1, the basic properties of m-to-DH type II bursts are given. In this table, the date of the event (yymmdd) is given in Column 1. Columns 2–6 give the m type II properties: starting and ending times of the events in UT in Columns 2 and 3, starting and ending frequencies of m type II radio bursts (in MHz) in Columns 4 and 5, and the m type II shock speed (as  listed in Culgoora online type II catalog) in column 6. Similarly, the properties of DH type II events are shown in Columns 7-10 (Columns 7 and 8 correspond to the starting and ending times of DH type II events in UT, and Columns 9 and 10 give the starting and ending frequencies of DH type II radio bursts in kHz). Column 11 defines whether type II bursts-associated-shocks and/or flares produced SEPs (Yes), or not (No). Please note that for few events, DH type II started slightly earlier than the end time of the m type II bursts. For an example, the 990604 event had  m type II burst's ending time at 07:10 in 30 MHz, but DH type II emission started little earlier at 07:05 in 14 MHz. This is possibly due to the complexity of the type II emissions in these events where the precise determination of m type II end time (and/or a precise DH type II start time) is very difficult. The several-minute advance of the DH type II emission could therefore be interpreted as an uncertainty of the start/end time determination.\\
\indent In Table 2, we compile the data on the soft X-ray flares and CMEs which are associated with m-to-DH type II bursts. First column gives the date of events; Columns 2–6 describe the X-ray flare events: Columns 2–4 give the start, end and peak times of the flare events in UT, respectively; Column 5 shows the locations of flares and flare peak flux is presented in Column 6. The X-ray flare is classified as A, B, C, M and X according to the peak flux intensity $<$10$^{-7}$, 10$^{-7}$ -- 10$^{-6}$, 10$^{-6}$ -- 10$^{-5}$, 10$^{-5}$ -- 10$^{-4}$ and $\geq$10$^{-4}$ W m$^{-2}$, respectively. The numerical value after the class refers to the multiplicity factor. Columns 7–12 give the characteristics of the CMEs; Columns 7 and 8 are the onset and times of first appearance of CME in LASCO FOV in UT; central position angle and width are given in Columns 9 and 10. Columns 11 and 12 represent the CMEs mean sky-plane speed and residual acceleration, respectively.\\
The statistical values (mean and median) of the parameters describing properties of CME, soft X-ray flares, delay times and m and DH type II bursts separately for SEP and non-SEP events are presented in Table 3. In Column 1 the names of phenomena are given. Column 2 denotes the properties of the considered events. Columns 3-4 and 5-6 give the estimated mean and median of the considered parameters for SEP-associated and non-SEP-associated events, respectively. The type II formation heights listed in the table are obtained from the back extrapolation method, addressed in Section 3.5.1. In the next section we present statistical analyses of these two sets of events.\\
\section{Results and discussion}
\subsection{\bf{Characteristics of SEP-associated and non-SEP-associated CMEs}}
\indent Kinematics and width are important properties of CMEs for understanding their production of SEP events. An extensive investigation of various CME speeds (sky-plane speed, projection corrected speed, and peak speed of the sky-plane), acceleration, and width is presented in this section.\\
\indent The sky-plane speeds are the average speeds of the CMEs within the LASCO FOV obtained by using the linear fit to the height–time measurements. But sky-plane speed is only the projected speed of the CMEs. From the LASCO coronagraph images we can only detect apparent speeds and widths of CMEs. The three-dimensional structure and actual speeds of CMEs remain unknown due to the projection effect, especially for disk CMEs (source longitude $\leq$ 60$^\circ$). Considering a set of energetic CMEs, \cite*{Gopal2000} demonstrated that the projection effect of speeds of the halo CMEs depends on the longitude \citep{Pappa2010,Gopal2007}. The projection corrected speed (real propagation speed) and width of halo CMEs can be obtained, among several methods, from a cone model \citep{Michalek2003,Xie2004}, it is deduced from the sky-plane speed by using the approximation method. We adopted the arguments of \cite*{Gopal2009b} to assign an average half width $(\omega)$ to each speed range and took it as the cone half angle: 66$^\circ$ $(V_{sky}$ $>$ 900 km s$^{-1})$, 45$^\circ$  (500 km s$^{-1}$ $<$ $V_{sky}$ $\leq$ 900 km s$^{-1}$), and 32$^\circ$ ($V_{sky}$ $<$ 500 km s$^{-1}$). The CME corrected speeds were calculated by using the relation $V_{corrected}=(cos\omega+sin\omega)/(cos\omega cos\theta +sin\omega)\times V_{sky}$ where $\theta$ is the angle between the cone axis and sky-plane obtained from the heliographic coordinates of the source. From this equation, we estimated the corrected speeds of partial and full halo CMEs. The residual acceleration from the \emph{CDAW SOHO/LASCO} CME Catalog was also derived from the height-time measurement. It is called as residual acceleration because the acceleration due to gravity and the propelling forces must have declined significantly, while the deceleration due to drag becomes dominant at heights corresponding to the \emph{LASCO} FOV.
\begin{figure}[h]
\begin{center}
 \includegraphics[width=3.3in]{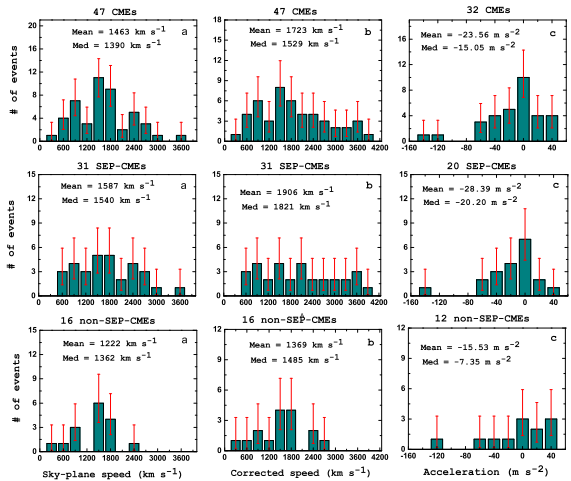}
\caption{ Distributions of: sky-plane speed (a); projection corrected speed (b); residual acceleration (c) of all events (top), SEP-associated CMEs (middle) and non-SEP-associated CME (bottom) events. The error bars denote the statistical error. }
\label{Fig3}
\end{center}
\end{figure}

\indent The distributions of sky-plane speed (a), corrected speed (b) and residual acceleration (c) of all events for which these properties could be determined (top) SEP-associated CMEs (middle) and non-SEP-associated CMEs (bottom) are shown in Figure 3. The difference between the mean sky-plane speed of SEP-associated CMEs and non-SEP-associated CME events (1587 km s$^{-1}$) and 1222 km s$^{-1}$), respectively) is statistically significant (student t-test shows there is P = 5\% chance that the difference is accidental). Prakash et al. (2010) also found that a set of CMEs associated with the m-to-DH type II radio bursts have the mean sky-plane speeds equal to 1655 km s$^{-1}$. However, the mean sky-plane speed of all CMEs (1463 km s$^{-1}$) is nearly three times greater than average CMEs \citep{Gopal2001a,Prakash2012a}. The error bars plotted in each bin for all figures are given by the calculation with 84.13\% of confidence level for upper and lower limits for small number of counts. They are based on standard equations derived from the Poisson and binomial statistics \citep{Gehrel1986}.\\
\indent The distribution of the projection corrected speed is shown in Figure3b. The student t-test shows that the difference between the mean corrected speeds of SEP-associated CMEs and non-SEP-associated CME events (1906 and 1369 km s$^{-1}$)) is statistically significant (P =2 \%). Recently, \cite*{Gopal2012b} studied the properties of a set of 16 GLE (Ground Level Enhancement) events. They found that, the mean (median) projection corrected speed of this set of events was 2083 km s$^{-1}$ (1937 km s$^{-1}$), which is slightly larger than that of our mean value.\\
\indent The distributions of residual acceleration of all CMEs (top), SEP-associated CMEs (middle) and non-SEP-associated CMEs (bottom) are shown in Figure 3c. The residual acceleration of all CMEs varies from --160 m s$^{-2}$ to 35 m s$^{-2}$ (mean = --23.56 m s$^{-2}$). The number of events in residual acceleration panels is reduced due to the restriction of sample of events. To get reliable values of acceleration we limited our study to events having at least four height-time data points, \emph{i.e.} those events without asterisk in Column 12 of Table 2 \citep{Gopal2001a,Yashiro2004}. The mean difference between residual acceleration of SEP-associated CMEs and non-SEP-associated CME events (--28.39 m s$^{-2}$ and --15.53 m s$^{-2}$, respectively) is statistically insignificant (P=41\%). However, if we exclude the two outliers (a highly decelerated event from each group) of SEP-associated CMEs and non-SEP-associated CMEs, the mean values are reduced to --21.52 m s$^{-2}$ and --5.63 m s$^{-2}$, respectively and it is statistically quite significant (P=8\%). From this result, we infer that SEP-associated CMEs are three times more decelerated than the non-SEP-associated CMEs. This result is consistent with \cite*{Vrsnak2004} and \cite*{Belov2014} that CMEs having high initial speed are mostly decelerated. From the present study, the mean sky-plane, corrected speeds of SEP-associated CMEs are much larger than non-SEP-associated CMEs. Due to the high initial speeds of SEP-associated CME events, they are more decelerated than the non-SEP-associated CMEs.\\
\indent To determine the sky-plane peak speed and its corresponding height, initially we selected height-time data from the \emph{CDAW} CME list and then smoothed these data. In order to find the sky-plane peak speed, we took the time derivative between each two height-time data points. From these values, we picked out the maximum (peak) speed and their corresponding height for our further analysis. The distribution of the sky-plane peak speed heights (a) and peak speed (b) for set of all CMEs (top), SEP-associated CMEs (middle) and non-SEP-associated CMEs (bottom) are shown in Figure 4. The 10\% of uncertainties in the estimation of sky-plane peak speed and their height is unavoidable. This may be due to the projection effect of disk events and also the different time cadences of LASCO C2 and C3 coronagraphs \citep{Prakash2012a}.
\begin{figure}[h]
\begin{center}
 \includegraphics[width=3.3in]{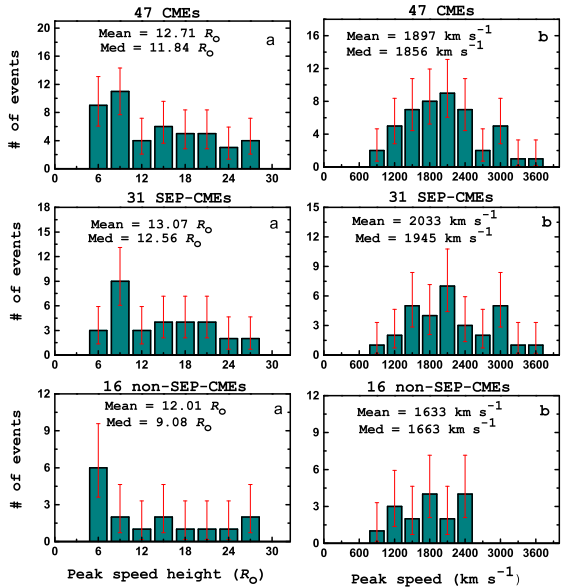}
\caption{Distributions of sky-plane: peak speed heights (a); peak speeds (b) of all CMEs (top) SEP-associated CMEs (middle) and non-SEP-associated CMEs (bottom). The error bars denote the statistical error.}
\label{Fig4}
\end{center}
\end{figure}
In this study, nearly half of m-to-DH type II radio bursts (all events) associated CMEs achieved their sky-plane peak speed within 12 \emph{R}$_\circ$ (median = 11.84 \emph{R}$_\circ$). The mean sky-plane peak speeds and their heights are distributed symmetrically. The t-test shows that difference between the average peak speed heights of SEP-associated CMEs and non-SEP-associated CME events (13.07 and 12.01 \emph{R}$_\circ$, respectively) is statistically insignificant. In both cases, CMEs reached their sky-plane peak speeds at nearly the same height. From this result, we reveal that the initial accelerating phase of all CMEs was continued in average up to 12 \emph{R}$_\circ$ (as seen from the Figure 4a-top). Recently, \cite*{Gopal2012b} and \cite*{Prakash2012b} also estimated the sky-plane peak speed height of a set of CMEs. They found that decelerating CMEs achieved their sky-plane peak speed at lower heights than the accelerating CMEs. Our result slightly differs from the other previous studies. In our case, more decelerated SEP-associated CMEs attained their peak speeds at larger heights (13.07 \emph{R}$_\circ$) than non-SEP-associated CMEs (12.01 \emph{R}$_\circ$).\\
\indent The distribution of the sky-plane peak speeds of CMEs associated with all CMEs (top), SEP-associated CMEs (middle) and non-SEP-associated CMEs (bottom) are shown in Figure 4b. The average peak speed of CMEs associated with all events (m-to-DH type II radio bursts) is 1897 km s$^{-1}$ (median = 1856 km s$^{-1}$). Interestingly within the average peak speeds height, the SEP-associated CMEs achieved higher speed (median=1945 km s-1) than that of non-SEP-associated CMEs (1663 km s$^{-1}$)). The difference between the mean sky-plane peak speed of SEP-associated CMEs and non-SEP-associated CMEs (2033 km s$^{-1}$ and 1633 km s$^{-1}$), respectively) is statistically significant (P=3\%).\\
The CME width seems to be a critical parameter to determining the kinetic energy of the CMEs \citep{Gopal2001a}. The DH type II bursts were also closely linked to CMEs which are faster and wider than average CMEs \citep{Gopal2001b}. The CME widths are one of the important parameter to decide SEP intensity \citep{Kahler1984}. Therefore, we studied the widths of the two populations of CMEs. However, it is difficult to assess the true width of halo CMEs. Note that the true width of halo CMEs cannot be determined from the 2D images because of the projection effect, especially for the disk event (longitude within $\pm$60$^\circ$). In our case, most of the CMEs originated from the disk center. Furthermore, the most of the SEP-associated CMEs (81\%) are full halo and only 56\% of non-SEP-associated CMEs are full halo. It should be noted that the mean apparent widths of the SEP-associated CMEs are slightly larger than the non-SEP-associated CMEs (330$^\circ$ and 274$^\circ$, respectively; please note that these distributions are not shown). Recently, \citep{Prakash2014} reported the widths of CMEs for a set of m-to-DH type II bursts associated geoeffetive CMEs. They found that the widths of the geoeffective CMEs (339$^\circ$) is considerably larger than non geoeffective CMEs (251$^\circ$).\\
\indent
\subsection{\bf{The CME speeds -- SEP peak intensity relationship}}
\begin{figure}[h]
\begin{center}
\includegraphics[width=3.3in]{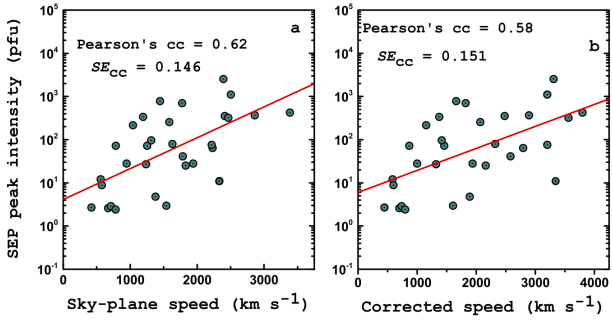}
\caption{The relationship between CME speeds and SEP peak intensity: sky-plane speed ? peak intensity (a); projection corrected speed ? peak intensity (b). The Pearson?s correlation coefficients cc and SEcc are indicated in the upper position. The straight line in red and the values are illustrate the linear fitting function and the corresponding Pearson statistical values.}
\label{Fig5}
\end{center}
\end{figure}
\indent The relationship between the sky-plane speed and corrected speeds of CMEs with peak intensity of the associated SEP events is shown in Figure 5. The speed of the CMEs should correlate reasonably well with the peak intensity of the associated SEP events \citep{Kahler1984,Reames1999,Kahler2001,Gopal2003} as one would expect when the particles are accelerated by the CME-driven shocks. Here, we also assume that protons are accelerated due to CME-driven shocks. In Figure 5, we found positive correlation between CME speeds (sky-plane speed and corrected speed) and the logarithmic peak intensity of SEP events. The estimated Pearson's correlation coefficients are \emph{cc} = 0.62 and \emph{cc} = 0.58, respectively. Although the SEP peak intensity is generally correlated with the CME speed, the scatter is very large. We also calculated the standard error of the correlation coefficient (\emph{SE}$_{cc}$) by using $SE_{cc}=\sqrt{(1-cc^2)/(n-2)}$, where n is the number of data points. The linear fitting functions of speed and intensity are indicated by the red straight lines. Recently, \cite*{Papaioannou2016} found that there is a good correlation between CME speed and proton peak flux (\emph{cc} = 0.57) for a different set of events and their standard error is around 0.07 (sample events are much larger than that of ours). From these studies, we infer that the speed of the CME associated with m-to-DH type II bursts may influence the peak intensity of the SEP events. Hence, these SEP-associated CMEs are very likely good electron and proton accelerators. Our results are consistent with previous studies (e.g., \citep{Gopal2005,Ding2013,SubraSha2016,Papaioannou2016}.\\
\subsection{Characteristics of SEP-associated and non-SEP-associated flares}
\indent The flare rise time (time interval between flare start and flare peak) is the most important parameter in analyzing the relationship between flares, CMEs and type II radio bursts \citep{Shan2003,Subramanian2003,Zhang2004,Prakash2009,Prakash2012c}. The distribution of rise times (a), duration (b), and peak flux (c) of all 45 events (top) SEP-associated flares (middle) and non-SEP-associated flare (bottom) events is shown in Figure 6. As seen from this figure, the average flare rise time of the 45 events with reported flares (see Table 2) is nearly equal to 28.51 min (median = 21 min). Most of the flares (71\%) reach the peak time within 30 min. There is no difference between the mean rise time of SEP-associated flares and that of non-SEP-associated flares (28.90 and 27.64 min, respectively). The t-test shows that the mean difference is statistically insignificant. Recently, \cite*{Prakash2012a} found that for eruptions associated with DH type II bursts, the median rise time of the flares associated with accelerating CMEs is larger than that in association with decelerating CMEs (median = 40.5 and 21.0 min).\\
\begin{figure}[h]
\begin{center}
\includegraphics[width=3.3in]{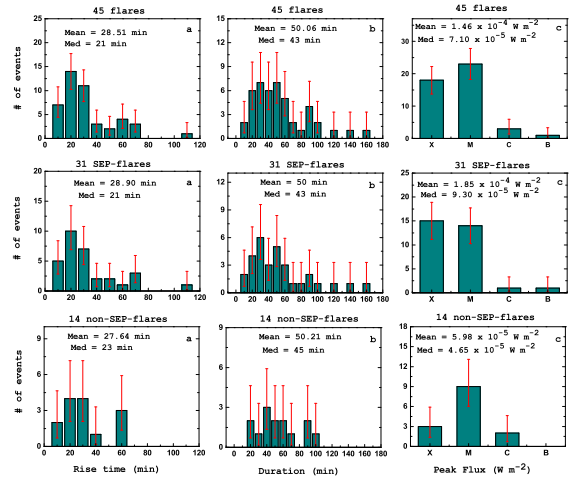}
\caption{Distributions of: rise times (a); duration (b); peak flux (c) of all (top), SEP-associated flares (middle) and non-SEP-associated flares (bottom). The error bars denote the statistical error.}
\label{Fig6}
\end{center}
\end{figure}
\indent The distribution of duration for 45 events (top), SEP-associated flares (middle) and non-SEP-associated flares (bottom) are shown in Figure 6b. For all m-to-DH type II radio bursts, associated flare duration (the interval from the start to end) varies from 10 – 160 min. The mean durations of both sets of events are nearly equal (50 min). Recently, \cite*{Prakash2010} found that the average duration of the flares associated with m-to-DH type II bursts was 54.24 min. Interestingly, the duration of SEP-associated flares is much lower than the flares associated with geoeffective events (74 min) events \citep{Prakash2014}. The propelling Lorentz force acts longer for geoeffective CMEs (it is consistent with the longer flare duration (or rise time)). But for SEP-associated CMEs the Lorentz force should be acting for a shorter time, so the Lorentz force and gravity decrease, the drag force becomes the dominant force because of the widths of the CMEs, and so they tend to decelerate more in the \emph{LASCO} FOV \citep{Vrsnak2004}.\\
\indent In Figure 6c, the distributions of peak flux of the 45 events with reported flares (top), SEP-associated flares (middle) and non-SEP-associated flares (bottom) are plotted. It should be noted that 48\% of SEP-associated flare and 21\% non-SEP-associated flare belongs to X class events, respectively. The differences in mean peak flux of SEP-associated flares and non-SEP-associated flares, 1.85 $\times$ 10$^{-4}$ W m$^{-2}$ and 5.98 $\times$ 10$^{-5}$ W m$^{-2}$, respectively is statistically significant (P = 5\%). The X-ray peak flux of SEP-associated flares (1.85 $\times$ 10$^{-4}$ W m$^{-2}$) is stronger than that of geoeffective storm associated flares (0.77 $\times$ 10$^{-4}$ W m$^{-2}$).\\
\subsection{\bf{Time delays between the properties of flares, CMEs and type IIs for SEP-associated and non-SEP-associated events}}
\indent We examined the time delays (relative timings) between properties of m-to-DH type II bursts, flares and CMEs for SEP-associated and non-SEP-associated events (see Table 3, note that these distributions are not shown). We have considered the delay times between the onset of CMEs (back-extrapolated time) and starting time of m and DH type II radio bursts. The mean time delays between the onset times of CMEs and starting time of m type II bursts for SEP-associated events is lower than that of non-SEP-associated events (10 and 16 min, respectively) and it is statistically insignificant. However, the difference of the average time delays between the onset times of CME and starting time of DH type II bursts for SEP-associated and non-SEP-associated events are 26 min and 34 min, respectively. It has quite low statistical significance (P = 10\%).\\
\indent Similarly, the time delays between flare onset and starting time of m type II burst ranges from 5 to 67 min. The 94\% of m type II associated with SEP events originated within the rising phase of the flare. But 78\% of non-SEP-associated events are formed before the peak of flare. However, there is no significant difference between the mean time delays between the flare onset and m type II bursts start for SEP-associated and non-SEP-associated events (19 min for both). On the contrary, the mean delay time between the flare onset and starting time of DH type II bursts for SEP events is lower than the time delays of non-SEP-associated events (34 and 40 min, respectively). The 52\% and 28\% of the SEP-associated and non-SEP-associated DH type II emission originated during rising phase of the flares. The average delay time between the beginning of the m and DH type II burst of SEP-associated events is slightly lower than that of non-SEP-associated events (16 and 21 min, respectively). From these results, we infer that the longer flare – type II delay may be related to the possibility that the shock forms at larger distance where the CME becomes super-Alfvénic \citep{Gopal2001b}.\\
\subsection{\bf{Characteristics of m and DH type II for SEP-associated and non-SEP-associated events}}
\indent In this section, we have performed similar analysis to study the characteristics of m and DH type II bursts for both sets of events. The drift rate is determined from starting and ending frequencies, and the duration (\emph{T}$_{d}$) of m and DH type II bursts $[df/dt= (fs- fe)/Td]$, where subscripts s and e stand for the starting and ending fre-quencies of type II radio emission. The distributions of the duration (a) and shock speed (b) of m type II bursts for SEP-associated (top) and non-SEP-associated (bot-tom) events are presented in Figure 7. As seen from this figure, the duration of the m type II bursts for both sets of events ranges from 3 to 29 min. There is no significant difference between the median duration of m type II bursts for SEP-associated and non-SEP-associated events (8 and 10 min, respectively). Most of the m type II bursts have the duration around 10 min. Our result is consistent with previous results \citep{Shan2003,Prakash2010,Vasanth2011,Prakash2014}.\\
\indent The differences between the mean starting frequencies of m type II bursts for SEP-associated and non-SEP-associated events is statistically insignificant (86 and 76 MHz, respectively; please note that these distributions are not shown). However, 68\% of SEP-associated m type II bursts have the starting frequencies below 100 MHz, and the remaining 32\% of events have relatively higher starting frequency ($>$100 MHz). This result is different from that for a set of geoeffective CMEs associated with m type II bursts studied by \cite*{Prakash2014}.\\
\begin{figure}[h]
\begin{center}
\includegraphics[width=3.3in]{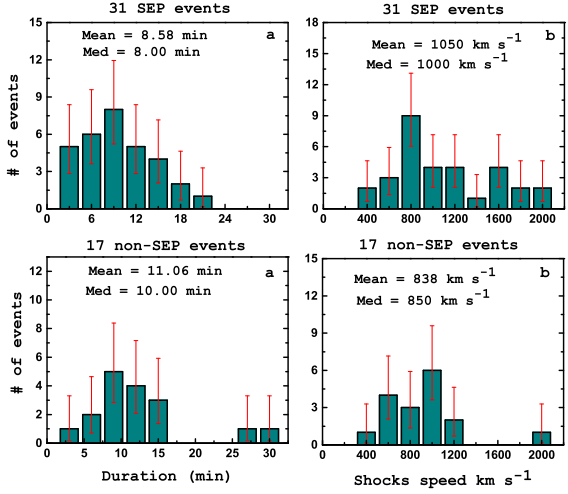}
\caption{Distributions of: duration (a); shock speed (b) of m type II radio bursts for SEP-associated (top) and non-SEP-associated (bottom) events. The error bars denote the statistical error.}
\label{Fig7}
\end{center}
\end{figure}
\indent They found that 62\% of m type II bursts associated with geoeffective CMEs have starting frequencies above 100 MHz, and the remaining 38\% of events have starting frequencies below 100 MHz. We also examined the difference between the ending frequencies of m type II bursts associated with both sets of events. There is no significant difference between the ending frequencies of m type II bursts for SEP-associated and non-SEP-associated events (25 and 23 MHz, respectively). However, the drift rate of m type II radio bursts associated with SEPs events is higher than for the non-SEP associated m type IIs. The difference between the mean drift rates of m type II radio bursts for SEP-associated and non-SEP-associated events (0.17 and 0.10 MHz s$^{-1}$) is statistically significant (P= 5\%). From this result, we infer that SEP-associated m type II shocks travelled with higher speeds than non-SEP-associated shocks.\\
\indent The distribution of shock speed derived from m type II radio bursts for SEP-associated (top) and non-SEP-associated (bottom) events is shown in Figure 7b. As seen from this figure, SEP-associated m type II shocks have higher speed (1050 km s$^{-1}$) than that of non-SEP-associated events (838 km s$^{-1}$). It is consistent with our previous result, the drift rate of the m type II bursts for SEP-associated (0.17 MHz s$^{-1}$) events are larger than non-SEP-associated events (0.10 MHz s$^{-1}$). The difference between the average shock speed of m type II burst for SEP-associated and non-SEP-associated events is quite statistically significant (P = 8\%).\\
\indent Figure 8 shows distributions of the duration (a) and ending frequency (b) of DH type II radio bursts for SEP-associated (top) and non-SEP-associated events (bottom). The durations of all DH type II bursts range from 17 min to 2650 min. However, most DH type II bursts have durations within 500 min. The peak of the distribution lies at 200 min for both sets of events. The difference between the mean durations (531 and 304 min, respectively) of the DH type II radio bursts for SEP-associated and non-SEP-associated events is statistically insignificant (P = 22\%).\\
\begin{figure}[h]
\begin{center}
\includegraphics[width=3.3in]{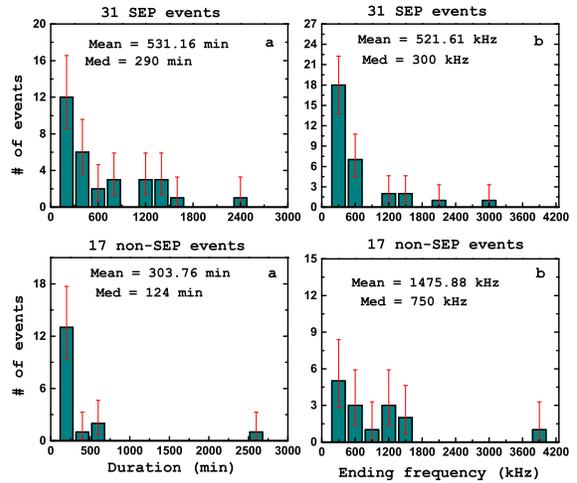}
\caption{Distributions of: duration (a); ending frequency (b) of DH type II radio bursts for SEP-associated (top) and non-SEP-associated (bottom) events. The error bars denote the statistical error. }
\label{Fig8}
\end{center}
\end{figure}

\indent The starting frequencies of both sets of events do not differ much because most of DH type II bursts have starting frequencies around the upper cut-off frequency of the WAVES instrument (14 MHz). The distribution of ending frequencies of DH type II bursts for SEP-associated (top) and non-SEP-associated (bottom) events is shown in Figure 8b. It is noted that 58\% of SEP-associated DH type II bursts enter into the km domain ($<$ 300 kHz). On the other hand, only 29\% (5/17) of non-SEP-associated events reach the km domain. The differences between the mean ending frequencies of DH type II bursts for SEP-associated and non-SEP-associated events (522 and 1476 kHz, respectively) are quite statistically significant (P = 6 \%). The ending frequency of type II bursts reflects the kinetic energy of CME \citep{Gopal2005}. It implies that the SEP-associated CMEs have larger kinetic energy than the non-SEP-associated CMEs.\\
\indent We also examined the drift rate of DH type II bursts for SEP-associated and non-SEP-associated events. We found that the average drift rate for SEP-associated DH type II bursts is lower than non-SEP-associated events (1.08 and 1.92 kHz s$^{-1}$, respectively). The t-test shows that the difference between the mean drift rate of these two populations is also statistically quite significant (P = 8 \%). It should be noted that, in comparison with non-SEP-associated type II bursts, SEP-associated type II bursts have higher drift rate in the m domain and lower drift rate in DH domain. This difference is in agreement with the fact that SEP-associated CMEs decelerate more than non-SEP-associated CMEs.\\
\subsection{\bf{Type II Formation height}}
~~ Based on delay time studies, the type II formation height is much closer to Sun for SEP-associated events and larger distance for non-SEP-associated events. In this section, we examined quantitatively the type II formation in coronal and interplanetary medium statistically. It is believed that type II radio bursts are generated by the shock formed around a CME moving at a speed faster than the local magnetosonic speed. It is still under debate where exactly the type II source locates at the shock front. For the metric type II source, some authors claimed that it is due to the shocks formed as a result of the expansion of the CME flanks \citep{Cho2008,Liu2009}, while some other authors believed that the shock ahead of the CME nose is a preferable place \citep{Ramesh2012}. For the DH type II source, different opinions also exist. For instance, \cite*{Magda2014} found that the type II sources were close to the flank for the CME they investigated. However, \cite*{Martnez2012} found that the shock ahead of the CME nose is more favorable for producing type II emission \citep{Shen2013}. In this section, the type II formation height was derived from two different methods. A straight-forward method is the inversion of the density corresponding to the starting frequency to a height using some analytical density model. However, the inverted heights may subject to large error bars due to the large uncertainties of density models. Therefore we also computed the type II formation height from the back extrapolation of the CME height to the height at the type II start time under the assumption of a constant CME speed. The values listed in Table 3 are from the back extrapolation method.\\
\indent A straight-forward method to estimate the type II formation height is to use the relationship between the plasma frequency fp and electron density ne given by $f_p=9 \times 10^{-3} \sqrt{n_{e}}$  MHz  the type II bursts observed at local plasma starting (f$_{s}$) and ending frequencies (\emph{f}$_{e}$) were used to get information about the plasma density (\emph{n}$_{e}$) during the starting and ending frequencies of type II emission. From these densities, one can estimate the two different heights (formation and ending height) of coronal shocks by using the one-fold Newkirk $n_e=4.2\times10^4\times10^{4.32(R_\circ/ R)}$ electron density model \citep{Newkirk1961}. We estimated the m type II formation height by using the starting frequencies; it varies in the ranges 1.45 – 1.86 \emph{R}$_\circ$. We also found the ending heights of m type II emission ranges from 1.73 to 3.15 \emph{R}$_\circ$. The difference in the mean value of formation height of m type II emission for SEP-associated and a non-SEP-associated event (1.41 and 1.64 \emph{R}$_\circ$, respectively) is statically insignificant.\\
\indent As the density model is subject to large uncertainties, we also checked type II formation height using the back extrapolation of the CME height at the type II start time under an assumption of a constant CME speed. As in \cite*{Gopal2009b}, another assumption we made is that the heliocentric distance of the m type II burst (and of the shock) can be approximated by that of the driving CME, more specifically, the leading edge distance. Consequently, the height can be simply derived from the observation of SOHO/LASCO using this relation $H_{type II}=R_{c}-V_{c}\times T_{cv}$, where \emph{R}$_{c}$ is the first appearance position of CMEs in LASCO C2/C3 field of view, \emph{T}$_{cv}$ is the time delays between the first appearance time of the CME and the start time of the m and DH type II burst. We assigned two different CME speeds (\emph{V}$_{c}$) to determine the formation height of m and DH type II radio bursts respectively. For a m type II burst, Vc is the initial speed estimated from the first two height-time data points within the LASCO FOV. For a DH type II burst, we utilized the projection corrected speed calculated from the sky-plane speed (as described in Section 3.1) within the full LASCO FOV. The constant speed method (linear speed assumption) has been used by several authors \citep{Harrison1986,Harrison1995,Moon2002,ZhangM2002,Andrews2003,Cho2003} to estimate the CME onset times, since we have insufficient kinematic information of CMEs at the time of type II radio bursts in many cases. The impulsive acceleration phases of the CMEs are coincident with the impulsive phases of GOES X-ray flares, and their accelerations nearly stop after the flare peak times \citep{Zhang2001,Neupert2001,Shan2003}. In our case, most of the type IIs started just before the flare peak times or after. So the speeds of flare-associated CMEs are nearly constant at the coronal region from 1.4 to 2.4 \emph{R}$_\circ$ \citep{MacFis1983}.\\
\begin{figure}[h]
\begin{center}
\includegraphics[width=3.3in]{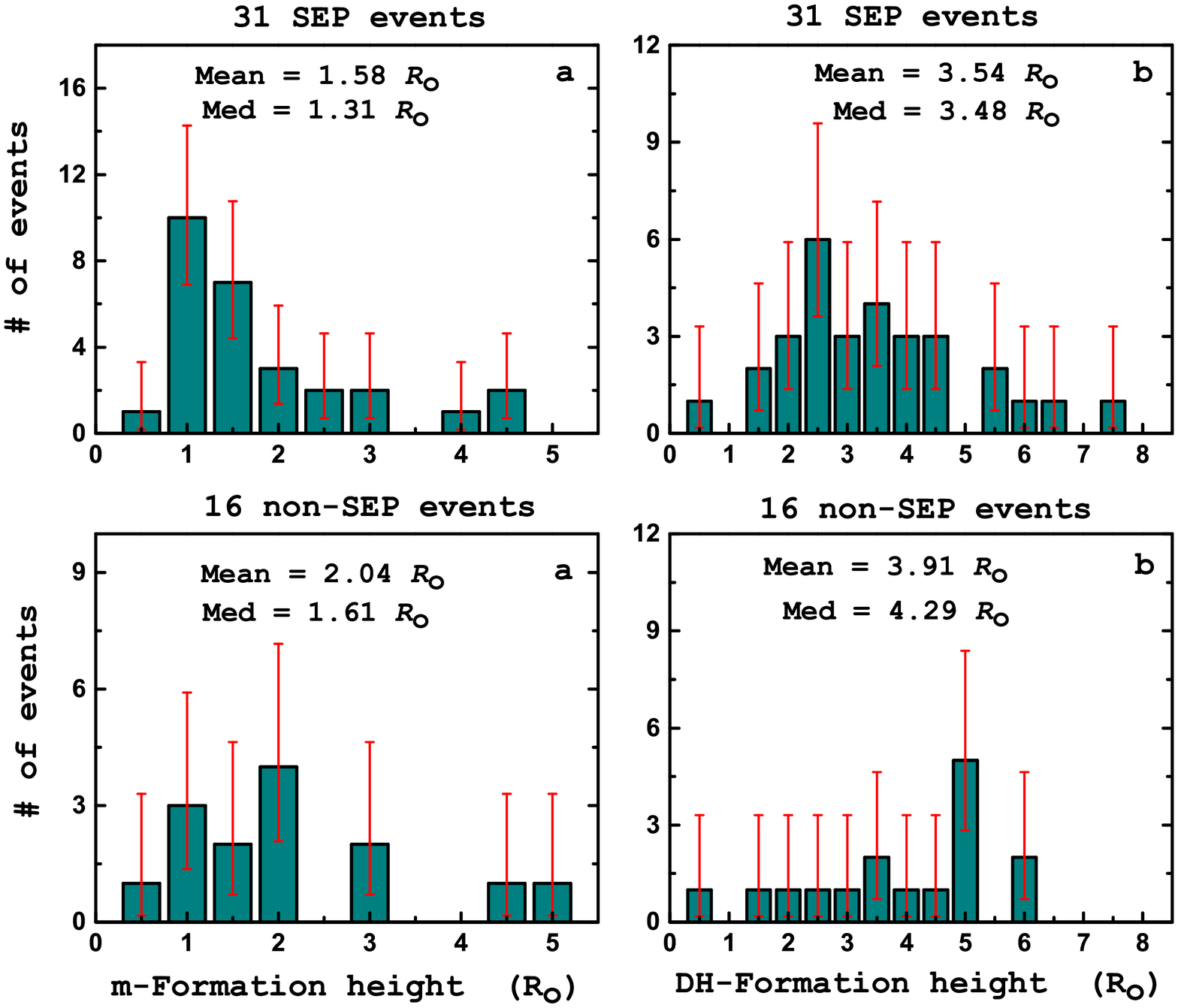}
\caption{Distributions of formation height of: m type II (a); and DH type II (b) for SEP-associated (top) and non-SEP-associated events (bottom) using the LASCO back extrapolation method. }
\label{Fig9}
\end{center}
\end{figure}

\indent These facts imply that the errors of the linear extrapolated heights at the time of type II radio bursts are less serious than in the case of CME onset time estimates. The formation height of m (a) and DH type II bursts (b) for SEP-associated (top) and non-SEP-associated (bottom) events are shown in Figure 9. As seen from this figure, 71\% and 43\% (SEP and non-SEP-associated events, respectively) of the formation height of m type II bursts lies within 1.5 Ro. The median formation heights of m type II bursts for SEP-associated and non-SEP-associated events (1.31 and 1.61 Ro, respectively) are nearly equal. The student t-test also shows that the difference between the mean values of these two populations (1.58 Ro and 2.04 \emph{R}$_\circ$) is statistically insignificant (P = 26\%). The median formation heights are well consistent with the estimated formation height using the Newkirk model previously. \cite*{Makela2015} analyzed the heights of a different population of CMEs at the onset of m type II bursts using the flare onset method. They obtained the mean m type II formation height is 1.72 \emph{R}$_\circ$ (median= 1.6 \emph{R}$_\circ$). Their results are slightly higher than the values in this paper. This could be partially due to the projection effect and/or the different methods of the formation height calculation.\\
\indent We also estimated the formation height of DH type II bursts for SEP-associated (top) and non-SEP-associated (bottom) events shown in Figure 9b. In this study, nearly 61\% and 75\% of DH type II bursts for SEP and non-SEP-associated events are formed after 2.5 \emph{R}$_\circ$. The t-test shows that difference between the mean DH type II formation height for SEP and non-SEP-associated events (3.54 and 3.91 \emph{R}$_\circ$) is statistically insignificant (P = 51\%). One possible interpretation of the insignificance is that quite a few events, either SEP or non-SEP, have a starting frequency close to the ionospheric cutoff frequency of radio emissions. Therefore, we do not see much difference of the formation height between SEP and non-SEP events.\\
\subsection{\bf{Source locations}}
\begin{figure}[h]
\begin{center}
\includegraphics[width=3.3in]{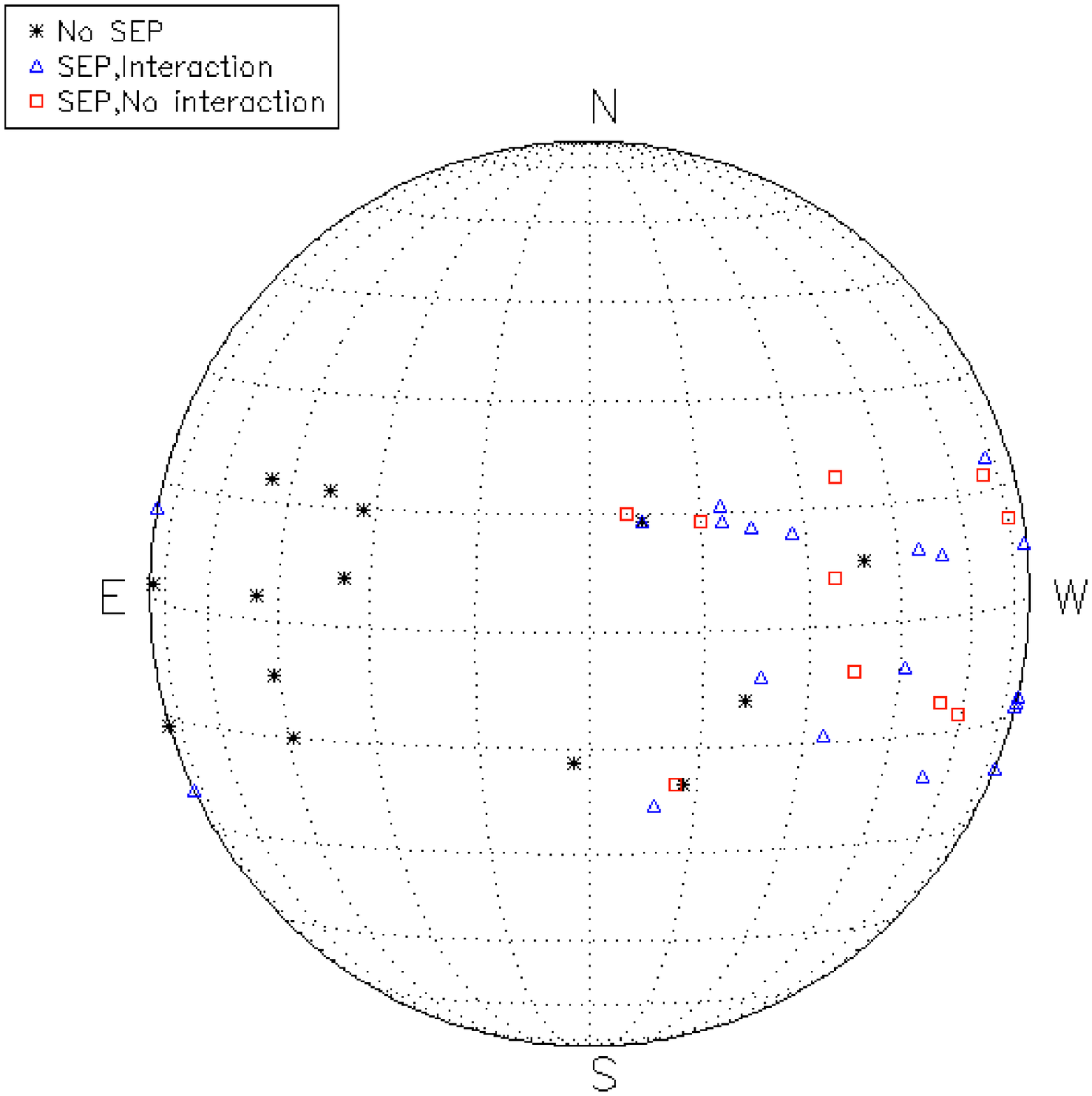}
\caption{Distributions of formation height of: m type II (a); and DH type II (b) for SEP-associated (top) and non-SEP-associated events (bottom) using the LASCO back extrapolation method. }
\label{Fig9}
\end{center}
\end{figure} 
\indent The distribution of the source location of 44 m-to-DH type II bursts events for SEP-associated and non-SEP-associated events are shown in Figure 10. From this figure, it should be noted that for all events, 73\% (32/44) originated from the disk center (longitude within ±60o). We also see that, 93\% of SEP-associated events are from the western side of the solar disk. On the other hand, 77\%of non-SEP-associated events originated from the eastern side. Furthermore, 67\% of the SEP-associated events sources lies within the disk center and the remaining 33\% of events from the limb, events are those with longitudes larger than 60$^\circ$. Out of 31 SEP-associated events, 20 (65\%) events are associated with CME-CME interaction and remaining 11 (35\%) events are non-interacting CMEs. The CME-CME interaction was examined by checking the position angle (PA) and time coincidence of primary SEP-associated CMEs and preceding one or more slow CMEs. It was also confirmed from the LASCO movies. The CMEs interaction is an important factor to decide whether CMEs are associated with SEP events or not \citep{Shan2014}. We also infer that the m-to-DH type II bursts associated solar flares and CMEs from the western hemisphere have a high possibility to be associated with SEPs that can be detected at Earth, also directly related with the connectivity of the magnetic spiral with Earth, is consistent with previous studies \citep{Wang2002,Gopal2002,Zhang2004,Zhang2007,Prakash2014,Vasanth2015}.\\
\indent We noticed that two events with their longitudes west of W15 did not produce SEPs as expected. For these two events, the shock speeds are lower than the average shock speed in the group of SEP- associated events. Probably the shock strength is not large enough to accelerate the $>$10 MeV protons to above 10 pfu which is our definition of a SEP event. Nevertheless, we checked the flux of some lower energy particles and they did show some enhancement related to these two events. On the other hand, we found that two events very close to the eastern limb were unexpectedly associated with SEPs. As these two events are interacting events, the propagation directions might be deflected after the primary CMEs erupted from their source regions. As a shock usually has a large extent in longitude, the accelerated SEPs may arrive the Earth.\\
\section{Summary and Conclusion}
\indent We have identified a set of m-to-DH type II bursts, solar flare and CMEs based on the Prakash et al. (2009), recorded during period of April 1997 – October 2012 (SC 23 and SC 24).  Out of 48 m-to-DH type II bursts associated with flares and CMEs, 31 (65\%) of them identified as SEP-associated events and the remaining 17 events could be interpreted as non-SEP-associated events. We analyzed the characteristics of m-to-DH type II bursts, CMEs, solar flares associated with SEP and non-SEP events. The main results of our study can be summarized as follows.
The mean sky-plane speed of SEP-associated CMEs is nearly three times as large as the mean sky-plane speed of average CMEs. There is a significant difference in the mean sky-plane speed of SEP-associated CMEs and non-SEP-associated CME events (1587 km s$^{-1}$ and 1222 km s$^{-1}$, respectively). The average projection corrected speed (real propagation speed) of SEP-associated CMEs is larger than that of the non-SEP-associated events (1906 and 1369 km s$^{-1}$). The SEP-associated CMEs found to be more decelerated than the non-SEP-associated CMEs within the LASCO FOV (--21.52 m s$^{-2}$ and --5.63 m s$^{-2}$, respectively). This implies that the mean residual acceleration of SEP-associated CME events is three times as great as the mean residual acceleration of the non-SEP-associated CMEs. The average apparent width of the SEP-associated CMEs is larger than the non-SEP-associated CMEs (330$^\circ$ and 274$^\circ$).\\
\indent The SEP-associated CMEs and non-SEP-associated CMEs achieved their sky-plane peak speed at nearly same heights (13.07 and 12.01 \emph{R}$_\circ$, respectively), but sky-plane peak speed of the SEP-associated CMEs (2033 km s$^{-1}$) is much larger than that of non-SEP-associated CMEs (1633 km s$^{-1}$). These results reflect that SEP-associated flares are stronger than the non-SEP-associated flares (1.85 $\times$ 10$^{-4}$ and 5.98 $\times$ 10$^{-5}$ W m$^{-2}$) even though the flare rise time (28.90 and 27.64 min) and duration (50 min) of both sets of events are nearly equal. As described earlier, the stronger flare implies that we have prolonged action of the Lorentz force \citep{Vrsnak2008}. We found positive Pearson's correlation between the speeds of CMEs (sky-plane speed and corrected speed) and logarithmic peak intensity of SEP events (\emph{cc} = 0.62 and \emph{cc} = 0.58, respectively). The delay time between m and DH type II bursts for SEP-associated events is smaller in average than non-SEP-associated events (16 and 21 min, respectively).\\
\indent There is no much difference between the duration, starting and ending frequencies of m type II bursts for SEP-associated and non-SEP-associated events. But the shock speed (1050 km s$^{-1}$ and 838 km s$^{-1}$) and drift rate (0.17 and 0.10 MHz s$^{-1}$) of m type II bursts for SEP-associated events are larger than the non-SEP-associated events. In case of DH type II burst, there is no difference between the starting frequencies of both sets of events. But the differences in the mean duration between SEP-associated and non-SEP-associated events (531 min and 157 min, respectively) is statistically high significant (P = 0.001\%). The 58\% of SEP-associated DH type II radio bursts enter into the km domain ($<$ 300 kHz, m-to-km type II bursts). On the other hand, only 28\% of non-SEP-associated DH type II events enter into the km domain. The mean difference between the ending frequencies (522 and 1476 kHz) and drift rates (1.08 and 1.92 kHz s$^{-1}$) of DH type II bursts for SEP-associated and a non-SEP-associated event is quite statistically significant. The average shock formation height of m (1.58 \emph{R}$_\circ$ and 2.04 \emph{R}$_\circ$) and DH type II bursts (3.54 \emph{R}$_\circ$ and 3.91 \emph{R}$_\circ$) for SEP-associated and non-SEP-associated event slightly differs. We found that 93\% of SEP-associated events from the westerns side of the solar disk and interestingly 65\% of SEP events are due the interaction of CMEs. From this study, at least for the set of CMEs associated with m-to-DH type II bursts, SEP-associated CME events are more energetic than those not associated with SEPs, thus indicating that they are effective particle accelerators.\\

\acknowledgments We thank the referee for useful constructive comments to improve the quality of this manuscript. We greatly acknowledge the data support provided by various online data centers of NOAA and NASA. We would like to thank the Wind/WAVES and Culgoora spectrograph teams for providing the type II catalogs. The SOHO/LASCO CME catalog is generated and maintained at the CDAW Data Center by NASA and The Catholic University of America in cooperation with the Naval Research Laboratory. SOHO is a project of international cooperation between ESA and NASA. This work is fully supported by NNSFC via grants 11233008, 11427803, 11473070, 11522328, and by MSTC via 2011CB811402. O. Prakash thanks to the Chinese Academy of Sciences for providing General Financial Grant from the China Postdoctoral Science Foundation. L. Feng also acknowledges the Youth Innovation Promotion Association and the research fund from the State Key Laboratory of Space Weather for financial support. G. Michalek was supported by NCN through the grant UMO-2013/09/B/ST9/00034.\\

\begin{table*}
\begin{center}
\small
\caption{Basic characteristics of m and DH type II bursts for SEP-associated and non-SEP-associated events\label{tbl-2}}
\begin{tabular}{@{}|c|ccccc|cccc|c|@{}}
\hline
    &\multicolumn{5}{c|}{m type II data}&\multicolumn{4}{c|}{DH type II data}& \\
\cline{2-6} \cline{7-10}
   \thead {Date\\(yymmdd)} &\thead{Start\\ Time\\(UT)}&\thead{End\\ Time\\(UT)}&\thead{Start\\ Freq\\ (MHz)}&\thead{End\\ Freq\\ (MHz)}&\thead{Shock\\ Speed\\(km s$^{-1}$)}&\thead{Start\\ Time\\ (UT)}&	\thead{End\\ Time\\ (UT)}&\thead{Start\\ Freq\\(kHz)}&\thead{End\\ Freq\\ (kHz)}& \thead{SEPs\\(Yes/No)} \\\hline

    971103&	0450&	0457&	45&	30&	500&	0515&	1200&	14000&	250&	No\\
    971104&	0558&	0607&	115&15&	1200&	0600&	0430(5)&	14000&	100&	Yes\\
980423&	0540&	0543&	70&	23&	1900&	0600&	1530&	14000&	200&	No\\
980509&	0326&	0329&	75&	23&	1500&	0335&	1000&	9000&	400&	Yes\\
990503&	0543&	0612&	90&	13&	450&	0550&	0845&	8000&	200&	No\\
990604&	0704&	0710&	90&	30&	800&	0705&	0100(5)&	14000&	60&	Yes\\
991116&	0246&	0256&	70&	35&	450&	0327&	0348&	7000&	4000&	No\\
991116&	0507&	0520&	45&	12&	950&	0517&	0534&	14000&	7000&	No\\
000710&	2121&	2133&	80&	28&	850&	2200&	2330&	14000&	1000&	No\\
000916&	0417&	0427&	70&	20&	1100&	0430&	1030&	14000&	400&	No\\
010120&	2118&	2126&	90&	20&	800&	2130&	2400&	14000&	500&	No\\
010402&	2152&	2157&	55&	28&	800&	2205&	0230&	14000&	250&	Yes\\
010410&	0513&	0517&	70&	23&	2000&	0524&	2400&	14000&	100&	Yes\\
010418&	0217&	0233&	130&	18X&	550&	0255&	1400&	1000&	100&	Yes\\
010512&	2340&	2348&	90&	20&	950&	2352&	0012&	3000&	1000&	No\\
011019&	0101&	0102&	110&	57X&	1800&	0115&	0225&	14000&	1300&	Yes\\
011226&	0512&	0516&	40&	26X&	800&	0520&	0500(27)&	14000&	150&	Yes\\
011228&	2018&	2037&	65&	26X&	400&	2035&	0300&	14000&	350&	Yes\\
020421&	0119&	0130&	65&	26X&	500&	0130&	2400&	10000&	60&	Yes\\
020720&	2110&	2120&	200&	55&	400&	2130&	2220&	10000&	2000&	Yes\\
020816&	0552&	0554&	90&	45&	1700&	0615&	0930&	14000&	300&	Yes\\
021027&	2300&	2310&	70&	18X&	900&	2306&	0120&	14000&	300&	No\\
030529&	0106&	0111&	85&	25&	800&	0110&	0800&	13000&	200&	Yes\\
031021&	0347&	0352&	50&	20&	1000&	0410&	0455&	5000&	1000&	No\\
031026&	0616&	0624&	160&	30&	1100&	0700&	0915&	8000&	1500&	No\\
040107&	0406&	0430&	55&	9X&	700&	0415&	0615&	14000&	750&	No\\
040623&	0601&	0608&	57&	30&	500&	0630&	0855&	14000&	5000&	No\\
040912&	0023&	0029&	57&	30&	900&	0043&	2100(13)&	14000&	40&	No\\
041030&	0614&	0624&	65&	20&	800&	0640&	0740&	4000&	1000&	Yes\\
041110&	0207&	0215&	130&	35&	1000&	0225&	0340&	14000&	1000&	Yes\\
050101&	0030&	0044&	100&	20&	800&	0045&	0225&	14000&	450&	No\\
050115&	2234&	2239&	45&	20&	1900&	2300&	0000&	3000&	40&	Yes\\
050709&	2203&	2210&	110& 35&	1000&	2215&	2300&	14000&	600&	Yes\\
050727&	0443&	0455&	60&	18&	700&	0520&	0645&	1000&	450&	Yes\\
050822&	0102&	0105&	35&	23&	1200&	0130&	0335&	8000&	550&	Yes\\
061213&	0227&	0235&	150&	15&	1600&	0245&	1040&	12000&	150&	Yes\\
061214&	2210&	2213&	70&	25&	1500&	2230&	2340&	14000&	1500&	Yes\\
110215&	0152&	0200&	130&	29X&	800&	0210&	0700&	16000&	400&	Yes\\
110607&	0626&	0640&	70&	15&	850&	0645&	1800&	16000&	250&	Yes\\
110802&	0608&	0622&	45&	12& 1000&	0615&	0730&	16000&	3000&	Yes\\
110804&	0353&	0410&	90&	18&	600&	0415&	1700(5)&	13000&	60&	Yes\\
110906&	0146&	0153&	150&	25&	1250&	0200&	2340&	14000&	200&	Yes\\
110906&	2219&	2234&	150&	20&	700&	2230&	1540(7)&	16000&	150&	Yes\\
120326&	2250&	2304&	95&	30&	400&	2315&	2355&	16000&	1500&	No\\
120517&	0132&	0142&	70&	13&	1100&	0140&	0620&	16000&	300&	Yes\\
120706& 2310&	2318&	45&	12&	1500&	2310&	0340(7)&	16000&	300&	Yes\\
120719&	0524&	0538&	50&	13&	1100&	0530&	0620&	5000&	600&	Yes\\
120927&	2344&	2352&	30&	18X	&700&	2355&	1015(28)&	16000&	250&	Yes\\
\hline

\end{tabular}
\tablecomments{In Column 5 the letter X indicates that the emission continued beyond the instrument range\\
In Column 8 the number in parentheses shows that the event extended in the following day}
\end{center}
\end{table*}

\begin{table*}
\begin{center}
\small
\caption{Basic characteristics of flares and CMEs associated with m-to-DH type II radio bursts\label{tbl-2}}
\begin{tabular}{@{}|c|ccccc|cccccc|@{}}
\hline
    &\multicolumn{5}{c|}{flare data}&\multicolumn{6}{c|}{CME data} \\
\cline{2-6} \cline{7-12}
   \thead {Date\\(yymmdd)} &\thead{Start\\ Time\\(UT)}&\thead{End\\ Time\\(UT)}&\thead{Peak\\ Time\\ (UT)}&\thead{Source\\Location}&\thead{Peak\\flux\\(W m$^{-2}$)}&\thead{CME\\Onset\\(UT)}&\thead{First\\Time\\ (UT)}&\thead{CPA\\(deg)}&\thead{Width\\(deg)}& \thead{Sky-plane\\Speed\\(km s$^{-1}$)}&\thead{A\\(m s$^{-2}$)}\\
\hline
971103&	0432&	0449&	0438&	S20W13&	C8.6&	0313&	0528&	240	&109&	227&	1.4*\\
971104&	0552&	0602&	0558&	S14W33&	X2.1&	0526&	0610&	Halo&	360&	785&	-22.1\\
980423&	0535&	0623&	0555&	S17E90&	X1.2&	0528&	0555&	Halo&	360&	1691&	-44.4*\\
990503&	0536&	0632&	0602&	N15E32&	M4.4&	0550&	0606&	Halo&	360&	1584&	15.8\\
980509&	0304&	0355&	0340&	S14W89&	M7.7&	0325&	0335&	262	&178&	2331&	-140.5*\\
990604&	0652&	0711&	0703&	N17W69&	M3.9&	0643&	0742&	289	&150&	2230&	-158.8\\
991116&	0236&	0254&	0246&	N17E38&	M3.8&	0213&	0306&	81&	98&	636&	-9.2\\
991116&	0447&	0545&	0512&	N08W39&	M1.8&	0509&	0530&	285	&129&	712&	22.4\\
000710&	2105&	2227&	2142&	N18E49&	M5.7&	2120&	2150&	67	&290&	1352&	35\\
000916&	0406&	0448&	0426&	N14W07&	M5.9&	0400&	0518&	Halo&	360&	1215&	-12.3\\
010120&	2106&	2132&	2120&	S07E46&	M7.7&	2108&	2130&	Halo&	360&	1507&	-41.1\\
010402&	2132&	2203&	2151&	N19W72&	X20.0&	2143&	2206&	261	&244&	2505&	108.5*\\
010410&	0506&	0542&	0526&	S23W09&	X2.3&	0521&	0530&	Halo&	360&	2411&	211.6*\\
010418&	0133&	0141&	0136&	S23W90&	B9.1&	0211&	0230&	Halo&	360&	2465&	-9.5\\
010512&	2242&	2345&	2335&	S17E02&	M3.0&	2318&	0245&	190	&132&	527&	3.1\\
011019&	0047&	0113&	0105&	N16W18&	X1.6&	0026&	0127&	Halo&	360&	558&	-25.6\\
011226&	0432&	0647&	0540&	N08W54&	M7.1&	0506&	0530&	281	&213&	1446&	-39.9*\\
011228&	2002&	2132&	2045&	S26E90&	X3.4&	2006&	2030&	Halo&	360&	2216&	6.9\\
020421&	0043&	0238&	0151&	S14W84&	X1.5&	0028&	0127&	Halo&	360&	2393&	-1.4\\
020720&	2104&	2154&	2130&	SE90b&	X3.3&	2049&	2218&	Halo&	360&	1941&	----\\
020816&	0546&	0631&	0611&	N07W83&	M2.4&	0553&	0606&	293	&162&	1378&	-3.7\\
021027&	----&	----&	----&	SE90b&	----&	2243&	2318&	Halo&	360	&2115&	75.1*\\
030529&	0051&	0112&	0105&	S06W37&	X1.2&	0046&	0127&	Halo&	360&	1237&	-22.3\\
031021&	----&	----&	----&	SE90b&	----&	0330&	0354&	Halo&	360&	1484&	-124.3\\
031026&	0557&	0733&	0654&	S15E44&	X1.2&	0613&	0654&	108&	208&	1371&	-62.1*\\
040107&	0343&	0421&	0404&	N02E82&	M4.5&	0347&	0406&	78	&171&	1581&	-60.4\\
040623&	0549&	0621&	0605&	S09W21&	C2.5&	----&	----&	----&	----&	----&	----\\
040912&	0004&	0133&	0056&	N03E49&	M4.8&	0031&	0036&	Halo&	360&	1328&	22.5\\
041030&	0608&	0622&	0618&	N13W22&	M4.2&	0608&	0654&	Halo&	360&	422&	25.9\\
041110&	0159&	0229&	0213&	N09W49&	X2.5&	0208&	0226&	Halo&	360&	3387&	-108.0*\\
050101&	0001&	0039&	0031&	N06E34&	X1.7&	0021&	0054&	Halo&	360&	832&	-5.5\\
050115&	2225&	2331&	2302&	N15W05&	X2.6&	2240&	2306&	Halo&	360&	2861&	-127.4*\\
050709&	2147&	2219&	2206&	N12W28&	M2.8&	2159&	2230&	Halo&	360&	1540&	-168.5*\\
050727&	0433&	0530&	0502&	N11E90&	M3.7&	0441&	0454&	Halo&	360&	1787&	-75.4\\
050822&	0044&	0218&	0133&	S11W54&	M2.6&	0101&	0131&	Halo&	360	&1194&	-17.8\\
061213&	0214&	0257&	0240&	S06W23&	X3.4&	0225&	0254&	Halo&	360&	1774&	-61.4\\
061214&	2107&	2226&	2215&	S06W46&	X1.5&	2200&	2230&	Halo&	360&	1042&	-0.4\\
110215&	0144&	0206&	0156&	S20W12&	X2.2&	0149&	0224&	Halo&	360&	669&	-18.3\\
110607&	0616&	0659&	0641&	S21W54&	M2.5&	0615&	0649&	Halo&	360&1255&	0.3\\
110802&	0519&	0648&	0619&	N14W15&	M1.4&	0557&	0636&	288	&268&	712&	-15.5*\\
110804&	0341&	0404&	0357&	N19W36&	M9.3&	0339&	0412&	Halo&	360&	1315&	-41.1\\
110906&	0135&	0205&	0150&	N14W07&	M5.3&	0200&	0224&	Halo&	360&	782&	105.6*\\
110906&	2212&	2224&	2220&	N14W18&	X2.1&	2158&	2305&	Halo&	360&	575&	1.1*\\
120326&	----&	----&	----&	BACK	&----&	2238&	2312&	Halo&	360&	1390&	-32.3\\
120517&	0125&	0214&	0147&	N11W76&	M5.1&	0125&	0148&	Halo&	360&	1582&	-51.8\\
120706&	2301&	2314&	2308&	S13W59&	X1.1&	2254&	2324&	Halo&	360&	1828&	-56.1\\
120719&	0417&	0656&	0558&	S13W88&	M7.7&	0513&	0524&	Halo&	360&1631&	-8\\
120927&	2336&	0034&	2357&	N06W34&	C3.7&	2331&	0012&	Halo&	360&947&	-27.1\\
\hline
\end{tabular}
\tablecomments{---- Means that there is no report of the corresponding data.\\
In column 5, 'SE90b' it means that the source is behind (b) the particular limb. Completely backside events with no information on the source location are marked as "BACK"\\
In Column 7, 'First time' denotes the time of CME at its first appearance in the LASCOC2/C3 field of view. CPA– central position angle; A– residual acceleration; *means that the value of acceleration is uncertain, otherwise it's reliable.}
\end{center}
\end{table*}

\begin{table*}
\begin{center}
\small
\caption{ Mean and median values of properties of CMEs, solar flares, relative timings, m and DH type II bursts for SEP-associated and non-SEP-associated events\label{tbl-2}}
\begin{tabular}{@{}|c|l|cc|cc|@{}}
\hline
   & &\multicolumn{2}{c|}{SEP-associated events}&\multicolumn{2}{c|}{Non-SEP-associated events} \\
\cline{3-4} \cline{5-6}
   Events &Properties&Mean&Median&Mean&Median \\
\hline
CMEs&Sky-plane speed (km s$^{-1}$)&	1587.09&	1540&	1222&	1361.50\\
	&Residual acceleration (m s$^{-2}$)&	-28.39&	-20.20&	-15.53&	-7.35\\
	&Corrected speed (km s$^{-1}$)&	1906.09&	1821.44&	1369.13&	1484.92\\
	&Peak speed height (\emph{R}$_\circ$)&	13.07&	12.56&	12.01&	9.89\\
	&Sky-plane peak speed (km s$^{-1}$)&	2033.21&	1944.93&	1633.08&	1662.67\\
	&Width (deg)&	329.50&	360.00&	273.50&	360.00\\
\hline
Solar flares&	Peak flux (W m$^{-2}$)&	X1.85&	M9.30&	M5.98&	M4.65\\
	&Duration (min)&	50.00&	43.00&	50.21&	45.00\\
	&Rise time (min)&	28.90&	21.00&	27.64&	23.00\\
\hline
Delay times&	CME onset-m type II delay (min)&	10.19&	9.00&	15.75&	12.00\\
	&CME onset-DH type II delay (min)&	25.74&	24.00&	33.94&	31.00\\
	&Flare onset-m type II delay (min)&	18.55&	12.00&	19.21&	17.00\\
	&Flare onset-DH type II delay (min)&	33.77&	29.00&	39.64&	40.00\\
	&m type II-CME first app delay (min)&	23.19&	20.00&	33.00&	22.50\\
	&DH type II –CME first app delay (min)&	11.42&	9.00&	12.94&	-1.50\\
	&m-DH type II delay (min)&	15.55&	14.00&	20.71&	20.00\\
\hline
m type II&	Duration (min)&	8.58&	8.00&	11.06&	10.00\\
	&Starting frequency (MHz)&	86.61&	70.00&	76.12&	70.00\\
	&Drift rate(MHz s$^{-1}$)&	0.17&	0.12&	0.10&	0.08\\
	&Shocks speed (km s$^{-1}$)&	1050&	1000&	838.24&	850\\
	&Formation height (Ro)&	1.58&	1.31&	2.04&	1.61\\
\hline
DH type II&	Duration (min)&	531.16&	290&	303.76&	124\\
	&Ending frequency (kHz)&	521.61&	300&	1475.88&	750\\
	&Drift rate(kHz s$^{-1}$)&	1.08&	0.59&	1.92&	1.50\\
	&Formation height (\emph{R}$_\circ$)&	3.54&	3.48&	3.91&	4.29\\
\hline
\end{tabular}
\end{center}
\end{table*}
\end{document}